\documentclass[a4paper,11pt]{article}
\pdfoutput=1 

\usepackage{jcappub} 

\usepackage[T1]{fontenc} 
\usepackage{aas_macros}
\usepackage{braket}
\usepackage{url}
\usepackage{tikz}
\usetikzlibrary{shapes,arrows}
\usepackage{subfigure}

\title{\boldmath Fast and accurate Gaia-unWISE quasar mock catalogs from LPT and Eulerian bias}

\author[a,b,c,d]{F. Sinigaglia,}
\author[c,d]{F.-S. Kitaura,}
\author[e]{M. Shiferaw,}
\author[c,d]{G. Favole,}
\author[e]{K. Storey-Fisher,}
\author[f,g]{N. Arsenov}

\affiliation[a]{Département d’Astronomie, Université de Genève, Chemin Pegasi 51, CH-1290 Versoix, Switzerland}
\affiliation[b]{Institut für Astrophysik, Universität Zürich, Winterthurerstrasse 190, CH-8057 Zürich, Switzerland}
\affiliation[c]{Instituto de Astrof\'isica de Canarias, Calle via L\'actea s/n, E-38205, La  Laguna, Tenerife, Spain}
\affiliation[d]{Departamento  de  Astrof\'isica, Universidad de La Laguna,  E-38206, La Laguna, Tenerife, Spain}
\affiliation[e]{Kavli Institute for Particle Astrophysics and Cosmology, Stanford University, 452 Lomita Mall, Stanford, CA 94305, USA}
\affiliation[f]{Institute of Astronomy and NAO, Bulgarian Academy of Sciences, 72 Tsarigradsko Chaussee Blvd., 1784 Sofia, Bulgaria}
\affiliation[g]{MTA--CSFK \emph{Lend\"ulet} ``Momentum'' Large-Scale Structure (LSS) Research Group, Konkoly Thege Mikl\'os \'ut 15-17, H-1121 Budapest, Hungary}

\emailAdd{francesco.sinigaglia@unige.ch}

\abstract{We present $100$ full-sky quasar spectrophotometric mock catalogs with smooth redshift evolution from $z=0$ to $z\sim 4$, tailored to analyze the Gaia-unWISE Quasar Catalog (\textit{Quaia}). In particular, we apply a novel hierarchical nonlocal nonlinear bias scheme (\texttt{Hicobian}) to dark matter fields generated through Augmented Lagrangian Perturbation Theory on the lightcone (\texttt{WebON} code), calibrating the free parameters of the bias model on \texttt{Abacus} quasar HOD mock catalogs tuned to reproduce DESI Early Data Release observations in real and redshift space. After having obtained such accurate spectroscopic catalogs, we inject in the mocks the observational effects characterizing the Quaia catalog: (i) spectrophotometric redshift uncertainties, (ii) the angular selection function, and (iii) the redshift number counts distribution. We assess the accuracy of our catalogs by validating a number of summary statistics: the full-sky QSO maps, the redshift uncertainty distributions as a function of redshift, the redshift $n(z)$ distribution, the angular power spectra and their normalized covariance matrices, and the angular two-point correlation functions. We find excellent agreement between these metrics from the mocks and from the Quaia catalog. We publicly release the mock catalogs to the community.}

\begin{document}
\maketitle
\flushbottom

\section{Introduction}
\label{sec:intro}

In the current era of precision cosmology, grand-scale state-of-the-art observational campaigns are delivering unprecedentedly wide, large and deep photometric and spectroscopic data sets. In particular, spectroscopic surveys such as DESI \cite{Levi2013}, Euclid \cite{Amendola2018}, J-PAS \cite{Benitez2014}, 4MOST \cite{Richard2019} and the future surveys conducted with the Nancy Grace Roman space telescope \cite[see e.g.,][]{Wang2022}, as well as photometric surveys such as LSST \citep{Ivezic2019} and DES \cite{DES2005}, promise to shed light onto key unsolved questions in cosmology, such as the nature on dark matter and dark energy. 

Among the different cosmological tracers probed by such observational efforts, quasars (hereafter QSOs) represent an invaluable probe of the large scale structure \citep[see e.g.,][]{White2012,Song2016,DESI2024,Richarte2025}. While sparser than other galaxy tracers such as luminous red galaxies and emission-line galaxies, QSOs are intrinsically very bright objects, which allow them to be observed over a very wide redshift range. Furthermore, QSOs are powered by accreting material feeding active galactic nuclei. Therefore, they represent a special population of objects constituted by supermassive black holes accreting material from their surroundings, tracing complementary density regimes to the other galaxy tracers.  

To adequately mine the wealth of information encoded in the aforementioned rich data sets, generating accurate and precise mock catalogs reproducing the expected clustering of dark matter tracers has become instrumental to estimate covariance matrices, test the ability of existing pipelines to cope with systematics, and validate new analysis techniques, among other applications. 

While QSO mock catalogs can in principle be obtained from $N$-body simulations by populating haloes with galaxies, using either the halo occupation distribution \citep[HOD, see e.g.,][]{NeymanScott1952,Peebles1974,Benson2000,BerlindWeinberg2002,Zheng2005,Zheng2007,Zentner2014,Hearin2016,Rocher2023} or subhalo abundance matching \citep[SHAM, see e.g.,][]{Vale2004,Conroy2006,Behroozi2010,TrujilloGomez2011,Reddick2013,Contreras2021,Favole2022,DeRose2022}, the computational burden behind running hundreds of different cosmic realizations makes it unfeasible to rely on such a procedure for the massive generation of mock catalogs. Instead, a wide range of approximate methods have been proposed in the literature, typically combining approximate gravity solvers with a variety of recipes to paint cosmological tracers onto the dark matter field \citep[e.g.,][]{Monaco2002,Kitaura2014,Avila2015,Kitaura2015,Chuang2015,Balaguera2019,Balaguera2020,Zhao2021,ColomaNadal2024}. 

In this paper, we present an effort aimed to generate mock catalogs mimicking the \textit{Gaia-unWISE} quasar catalog \citep[hereafter \textit{Quaia},][]{StoreyFisher2024}. Specifically, while attempting at generating mock catalogs for Quaia, a practitioner faces several different problems. First, the Quaia catalog covers the entire sky and goes out to redshift $z\sim 4$. This implies that a Quaia mock catalog must cover a cosmological volume of size $V=(10  \, h^{-1} \, {\rm Gpc})^3$, unless box replication is adopted. Second, given the large redshift interval probed by the Quaia catalog, the dark matter field used to paint the quasar sample onto must contain proper redshift evolution. This aspect is important for several reasons. A correct redshift evolution in the underlying dark matter field allows us to properly model the redshift evolution of the bias, and hence, of the cosmological tracers. This guarantees to have the correct redshift evolution of the relevant summary statistics, both traditional n-point and alternative statistics (such as e.g. cluster and void statistics, as already studied in these mocks by \cite{Arsenov2026}). In addition, the underlying dark matter lightcones could be used to produce multitracer mocks beyond Quaia QSOs, such as mocks for galaxy clustering, weak lensing, and the Lyman-$\alpha$ forest analyses, which require the right redshift evolution as opposed to a single redshift snapshot. Third, one needs to find the proper effective connection between the dark matter field and the quasar sample. This implies having accurate reference QSO catalogs available, produced either via HOD or SHAM applied to full $N$-body simulations. Fourth, the tradeoff between large volume and memory requirements imposes the usage of a coarse mesh ($l\sim 5.5 \, h^{-1} \, {\rm Mpc}$ side, in this case). To accomplish this but still have reliable small-scale clustering, an appropriate subgrid model to correctly populate the mesh with QSO positions inside each cell must be adopted.

Here, we present the end-to-end methodology that we employ to generate Quaia mock catalogs and validate them. In particular, we first lay out the procedure that we follow to produce `clean' (i.e. with no observational systematics) quasar spectroscopic catalogs. Subsequently, we inject Quaia observational effects into the spectroscopic catalogs. Finally, we validate the resulting clustering of the mocks by performing a comparison with the measurements from the true Quaia catalog.

The paper is structured as follows. In \S\ref{sec:quaiacat} we briefly summarize the main features of the Quaia catalog. In \S\ref{sec:flow} we lay out the full methodology that we adopt throughout the work. \S\ref{sec:validation} validates the mock catalogs by comparing a variety of summary statistics to the ones extracted from the Quaia catalog.\S\ref{sec:systematics} addresses the impact of observational systematics using the mock catalogs produced herein. We conclude in \S\ref{sec:conclusions}.    


\section{The Quaia catalog}\label{sec:quaiacat}

The Quaia QSO catalog \citep[][]{StoreyFisher2024} is a full-sky quasar catalog built combining public data from the \textit{Gaia} \citep{Delchambre2023,Gaia2023a,Gaia2023b} and \textit{WISE} \cite{Wright2010} surveys. Gaia DR3 includes a sample of $6,649,162$ quasar candidates, serendipitously detected during observations. Thanks to Gaia's low-resolution spectrograph covering the wavelength range $330-1050$ nm, these sources have redshift estimates, with $\sim 53\%$ of such sources having $|\Delta z/(1+z)|<0.01$ compared to SDSS redshift, i.e. at much higher precision than photometric redshifts. The sample defined this way has a median redshift $z=1.67$. The purity of this catalog is $\sim 52\%$, given that Gaia quasar target identification preferred completeness over purity.   

To reduce the fraction of contaminants and increase the purity of the sample, as well as to decrease the fraction of catastrophic redshift outliers due to line misidentification, the optical Gaia photometry was combined with \textit{unWISE} infrared photometry \cite{Lang2014} (a reprocessing of WISE data). In particular, the issue of purity was addressed by performing color and proper motion cuts. To improve redshifts accuracy and precision, SDSS quasars counterparts of Gaia's observations are used (when they exist) to train a $k$-Nearest-Neighbour ($k$NN) based on the photometry and Gaia-estimated redshifts. 

Finally, to alleviate the systematics imprinted by observational effects such as Galactic dust, a model for the selection function was fitted using Gaussian processes.

The final catalogs obtained with G-magnitude cuts $G<20.5$ ($G<20$) contain $1,295,502$ ($755,850$) sources, effectively covering $\sim64\%$ of the sky once the selection function is taken into account. Furthermore, the application of the $k$NN algorithm improves significantly the quality of Quaia's redshifts, with $\sim94\%$ ($\sim75\%$) spectrophometric redshifts in the $G<20.0$ catalog having precision $|\Delta z/(1+z)|<0.2 \, (0.01)$ with respect to the counterparts observed in SDSS. 

We refer the reader to the Quaia data release paper \cite{StoreyFisher2024} for further details about the construction of the catalog. The catalog is publicly available at: \url{https://doi.org/10.5281/zenodo.10403370}.


\section{Mock generation workflow} \label{sec:flow}

In this section, we present the workflow of our mock generation methodology. Fig. \ref{fig:scheme} shows a schematic representation of all the building blocks and the logical connections which constitute our methodology. Specifically, a gravity solver (\texttt{ALPT}, see \S\ref{sec:lightcones}) is applied to the initial conditions and outputs the dark matter particle positions and velocities, as well as the dark matter density field obtained by interpolating the DM particles onto a grid by means of a Cloud-in-Cell \citep[CIC,][]{HockneyEastwood1981} mass assignment scheme including phase-space mapping \citep{Abel2012}. With these items, we then produce (i) QSO number counts by applying a bias scheme to the DM field, (ii) real-space QSO positions, by combining the QSO number counts in cell and a suitable subgrid model, and (iii) redshift-space QSO positions, by displacing the QSO positions in real space by applying a model for redshft space distortions. All these operations are described in detail in what follows. We note that the same scheme is applied to both the `calibration' and the `production' phase: after the calibration has finalized and the best-fit values for the free parameters have been found, we apply the same procedure to the new independent cosmic realizations.

\tikzstyle{decision} = [diamond, draw, fill=blue!20, 
text width=6em, text badly centered, node distance=3cm, inner sep=0pt]
\tikzstyle{block} = [rectangle, draw, fill=blue!20, 
text width=9em, text centered, rounded corners, minimum height=4em, node distance=2.5cm,]
\tikzstyle{mock} = [diamond, draw, fill=green!20, 
text width=9em, text centered, rounded corners, minimum height=4em, node distance=4cm,]
\tikzstyle{block2} = [rectangle, draw, fill=yellow!20, 
text width=9em, text centered, rounded corners, minimum height=4em]
\tikzstyle{line} = [draw, -latex']
\tikzstyle{cloud} = [draw, ellipse,fill=red!20, node distance=2.5cm,
minimum height=4em]
\tikzstyle{cloudleft} = [draw, ellipse,fill=red!20, node distance=4cm,
minimum height=4em]

\begin{figure}
  \centering
  \begin{tikzpicture}[node distance = 2cm, auto]
    \node [block] (ICs) {\Large Initial conditions};
    \node [cloud, below of = ICs] (alpt) {\Large ALPT};
    \path [line] (ICs) -- (alpt);
    \node [block, below of = alpt, yshift=0cm, xshift=-5.5cm] (dmref) {\large Approx. DM density field};
    \path [line] (alpt) -- (dmref);
    \node [cloud, below of = dmref] (bias) {\Large Bias};
    \path [line] (dmref)--(bias);
    \node [block, below of = alpt, yshift=0cm, xshift=0cm] (dmpos) {\large Approx. DM particles positions};
    \path [line] (alpt) -- (dmpos);
    \node [block, below of = alpt, yshift=0cm, xshift=+5.5cm] (dmvel) {\large Approx. DM particles velocities};
    \path [line] (alpt) -- (dmvel);
    \node [cloud, below of = dmpos] (subgrid) {\Large Subgrid model};
    \path [line] (dmpos) -- (subgrid);
    \node [cloud, below of = dmvel, yshift=0cm, xshift=0cm] (rsd) {\Large RSD};
    \path [line] (dmvel) -- (rsd);
    \node [block, below of = bias, yshift=0cm, xshift=0cm] (ncounts) {\large Approx. tracer number counts};
    \path [line] (bias) -- (ncounts);
    \node [block, below of = subgrid, yshift=0cm, xshift=0cm] (rspacepos) {\large Real space tracer positions};
    \path [line] (ncounts) -- (rspacepos);
    \path [line] (subgrid) -- (rspacepos);
    \node [block, below of = rsd, yshift=0cm, xshift=0cm] (zspacepos)
    {\large Redshift space tracer positions};
    \path [line] (rspacepos) -- (zspacepos);
    \path [line] (rsd) -- (zspacepos);
    \node [mock, below of = zspacepos, yshift=0cm, xshift=0cm] (mock)
    {\Large Mock catalog};
    \path [line] (zspacepos) -- (mock);
    \path [line] (rspacepos) -- (mock);
  \end{tikzpicture}
  \label{fig:scheme}
  \caption{Workflow of the mock generation method presented in this work.}
\end{figure}
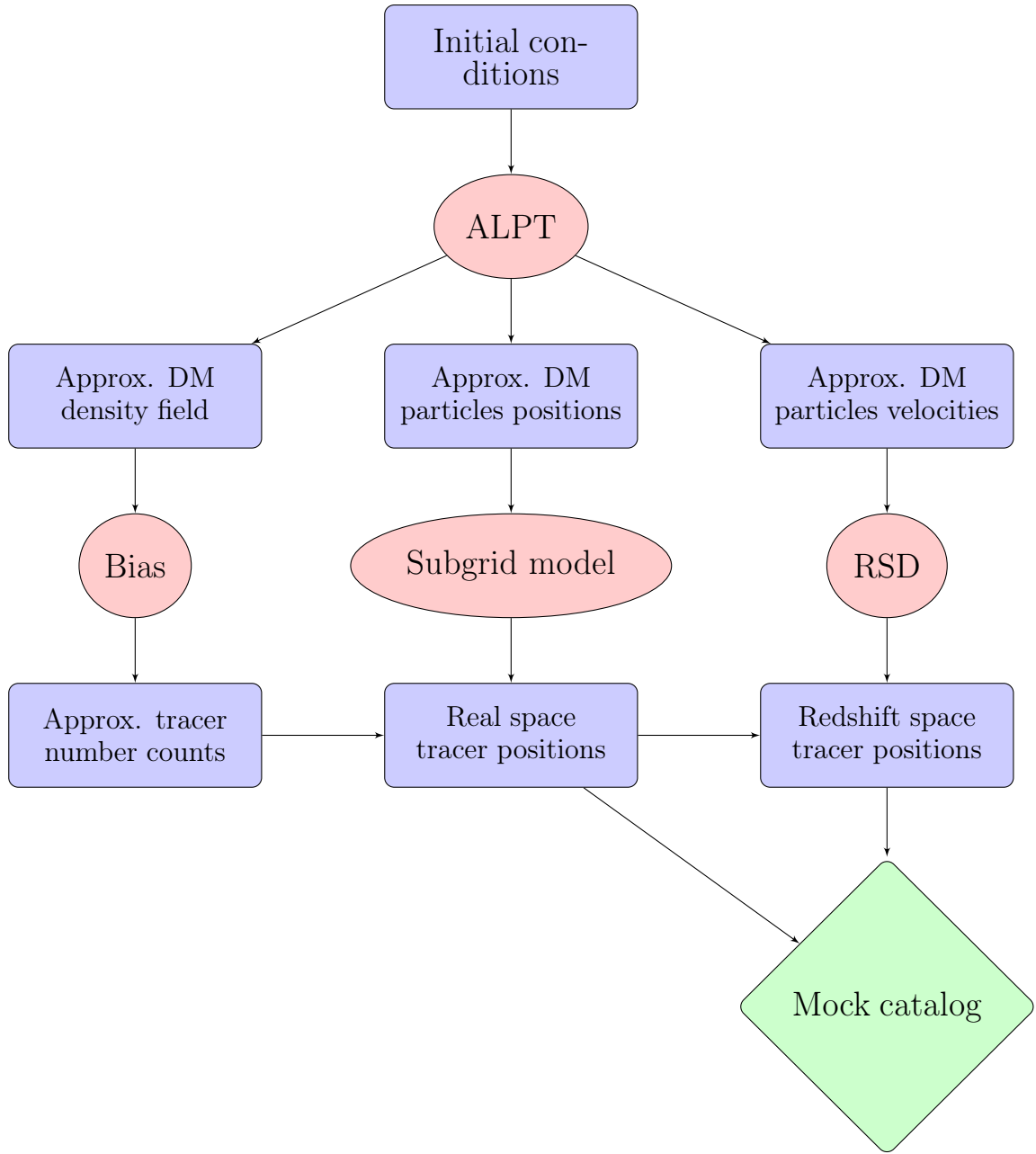

\subsection{Dark matter lightcones}
\label{sec:lightcones}

To generate the lightcone dark matter fields needed to produce the QSO mock catalogs, we rely on the \texttt{WebON} code (Kitaura \& Sinigaglia, in prep.), which is a C++ software implementing the \textit{(eulerian) Augmented Lagrangian Perturnation Theory} \citep[\texttt{(e)ALPT,}][]{Kitaura2013,Kitaura2024}. Briefly, in Lagrangian Perturbation Theory (LPT) the initial Lagrangian positions $\vec{q}$ are mapped onto the final Eulerian positions $\vec{x}$ through the displacement field $\vec{\Psi}$ as $\vec{x}=\vec{q} + \vec{\Psi}(\vec{q})$. Within the \texttt{ALPT} approximation, the displacement field is expressed as the combination of a long-range and a short-range component $\vec{\Psi}(\vec{q})=\vec{\Psi}_L(\vec{q})+\vec{\Psi}_S(\vec{q})$. The long-range displacement $\vec{\Psi}_L(\vec{q})$ is assumed to be the second-order LPT (2LPT) displacement $\vec{\Psi}_{\rm 2LPT}$ filtered through a Gaussian kernel $\mathcal{K}=\exp(-|\vec{q}|^2/(2r_s^2))$ regulated by a smoothing scale $r_s$: $\vec{\Psi}_L(\vec{q})=\vec{\Psi}_{\rm 2LPT}(\vec{q})*\mathcal{K}$. The short-range component $\vec{\Psi}_S(\vec{q})=\vec{\Psi}_{\rm SC}(\vec{q})*(1-\mathcal{K})$ is built upon the spherical collapse model $\vec{\Psi}_{\rm SC}(\vec{q})$ \cite{Bernardeau1994,Mohayaee2006}. 

The \texttt{ALPT} solution is implemented on the lightcones as follows:
\begin{itemize}
    \item we first compute the displacement field in the full box for the following set of fixed redshift snapshots: $z=\{0,~0.5,~1.0,~1.5,~2.0,~2.5,~3.0,~3.5\}$;
    \item we then loop over all the cells of the lightcone box, computing the distance from the observer position (placed at the center in this case) and the corresponding redshift;
    \item we displace the particle at the center of each cell, retrieving the displacement from the closest redshift snapshot to the Lagrangian redshift of the particle; 
    \item to ensure a smooth redshift evolution, we add a correction built upon the Zel'dovich approximation which is proportional to the difference of the growth factors at the redshift of the chosen snapshot and the `true' Lagrangian redshift \citep[see][]{AnguloWhite2010}. 
\end{itemize}

We refer the reader to Kitaura \& Sinigaglia (in prep.), for a thorough explanation of the details of the \texttt{WebON} code. 

We produced $100$ realizations with different random seeds of volume $V=(10 \, h^{-1} \, {\rm Gpc})^3$ on a $N_c=1800^3$ cell mesh, corresponding to a physical cell resolution $l\sim 5.5 \, h^{-1} \, {\rm Mpc}$. As anticipated, the observer is placed at the center of the box, meaning that the resulting lightcone DM fields cover the full sky out to $z\sim 4$. The runs were performed using the \texttt{Leonardo} pre-exascale supercomputer at CINECA (Italy), on computing nodes with $512$ GB RAM and $N=256$ cores. Each run took $\sim 12$ hours using the full node. 

\subsection{Reference QSO HOD catalogs}
\label{sec:ref_cat}

The `reference'\footnote{Throughout this work, we refer to the fiducial QSO catalog built by applying HOD onto N-body simulations and that we use to calibrate the bias model as `reference QSO catalog'.} QSO catalog used in this work to calibrate our bias model is built upon the publicly-available halo catalog from the \texttt{AbacusSummit} suite of simulations \citep{Maksimova2021}, based on the \texttt{Abacus} code \citep{Garrison2019, Garrison2021}. The suite comprises over $150$ simulations run with 97 different variations of cosmological parameters around the ``base'' Planck cosmology \citep{Planck2018}, performed with $6912^3$ particles in a $(2~\, h^{-1}\, {\rm Gpc})^3$ volume, corresponding to a mass resolution of $\sim 2.1\times 10^9 \, h^{-1} \, {\rm M}_\odot $.

In particular, we adopt a halo occupation distribution (HOD) approach, relying on the model by \cite{Zheng2007}, with five free parameters. The values for the free parameters are drawn from the publicly-available compilation by \cite{Yuan2024}. In particular, we use the \texttt{AbacusHOD} package \citep{Yuan2022} to apply the HOD to the Abacus snapshots, consistently with the procedure from \cite{Yuan2024}. This choice is motivated by the need of having a state-of-the-art simulation fitting the DESI QSO clustering to be able to calibrate our bias model onto. However, in principle the mock catalogs constructed in this way will reproduce the DESI QSO population, not the Quaia one. Therefore, in a second stage, as will be described later on in the paper, the bias model is fine-tuned to actually reproduce the summary statistics measured from the Quaia catalog.  
We notice that another possible approach would consist in fitting a HOD model directly to Quaia data. While this is principle feasible, it involves dealing with the complex observational systematics. This is not trivial as the systematics are intrinsically related to the lightcone geometry. A robust way to proceed would imply constructing a halo field on the lightcone and forward-model the observational systematics in the HOD fitting procedure, making it very expensive. As we will show in \S\ref{sec:systematics}, two-point clustering results are robust against these systematics, so one could in principle bypass the latter step. However, the impact of systematics is expected to become more important in 3D clustering analysis. For these reasons, this alternative methodology will be investigated in a future paper (Shiferaw et al., in prep.). Here, we prefer here to first fit the bias to the high-fidelity Abacus HOD catalogs to break as much as possible the parameters degeneracies, and then fine tune the parameters to reproduce the Quaia clustering.

\subsection{Bias model: theory}
\label{sec:bias_theory}

To populate our lightcone DM fields with QSO, we adopt the novel \texttt{Hicobian} nonlinear nonlocal stochastic bias model \citep{ColomaNadal2024}. In particular, \texttt{Hicobian} consists in a Eulerian parametric bias framework, which models nonlocal bias through a hierarchical cosmic web classification. In particular, \texttt{Hicobian} performs a first classification of each cell into four cosmic web environments \citep[known as $\mathcal{T}$-web,][]{Hahn2007} according to the relative values of the eigenvalues $\{\lambda_i, \,\, {\rm for} \, i=1,2,3; \, \lambda_1\ge\lambda_2\ge\lambda_3\}$ of the gravitational tidal field tensor $\mathcal{T}_{ij}=\partial_i\partial_j\phi$ with respect to a threshold $\lambda_{\rm th}$:
\begin{itemize}
    \item knots, if $\lambda_1,\lambda_2,\lambda_3\ge\lambda_{\rm th}$;
    \item filaments, if $\lambda_1,\lambda_2\ge\lambda_{\rm th},\lambda_3<\lambda_{\rm th}$;
    \item sheets, if $\lambda_1\ge\lambda_{\rm th},\lambda_2,\lambda_3<\lambda_{\rm th}$;
    \item voids, if $\lambda_1,\lambda_2,\lambda_3<\lambda_{\rm th}$.
\end{itemize}

Inspired by the $\mathcal{T}$-web classification and by previous studies hinting towards the importance of short-range nonlocal terms \citep[e.g.,][]{HeavensPeacock1988,McDonaldRoy2009,Sinigaglia2021,Sinigaglia2022,Kitaura2022}, \cite{ColomaNadal2024} proposed a further classification based on the relative values of the eigenvalues of the curvature tensor $\mathcal{D}_{ij}=\partial_i\partial_j\delta$ (let us dub it $\mathcal{D}$-web). In particular, within each $\mathcal{T}$-web environment, each cell is subclassified into short-range environments in the same way as above, according to a new threshold $\lambda^\prime_{\rm th}$. This hierarchical classification hence yields sixteen different environments, and cosmological tracers living in different environments can be treated as separate populations. In this way, the number of free parameters is effectively reduced to the number of free parameters necessary to describe just one environment. We refer the reader to \cite{ColomaNadal2024} for a thorough characterization of the hierarchical cosmic web classification.

It is worth stressing that such a phenomenological classification can be analytically linked to Eulerian perturbative expansions \citep[e.g.,][]{McDonaldRoy2009} and has been shown to model local and nonlocal terms up third order in such a framework \citep{Kitaura2022}, i.e. the relevant perturbative orders determining the two-point statistics of tracers \cite[e.g.,][]{WernerPorciani2020}.

Beside the nonlocality discussed above, the parametric bias implemented in \texttt{Hicobian} includes nonlinear, stochastic, and threshold bias. In our framework, the probability of finding a given number counts of objects $N_i$ in a cell $i$ of volume $dV$ is given by the following parametrization of the negative binomial distribution:
\begin{equation}
    P(N_i|\lambda_i,\beta)=\frac{\lambda_i^{N_i}}{N_i!}\exp(-\lambda_i) \times \frac{\Gamma(\beta+N_i)}{\Gamma(\beta)(\beta+\lambda_i)^{N_i}}\frac{\exp(\lambda_i)}{(1+\lambda_i/\beta)^\beta} \quad ,
\end{equation}
where $\lambda_i=\braket{\rho_{\rm tr}} \times dV$ is the expected number counts per cell and $\rho_{\rm tr}$ is the density of tracers, $P(N_i|\lambda_i,\beta)=\frac{\lambda_i^{N_i}}{N_i!}\exp(-\lambda_i)$ is the Poisson distribution modelling the statistical discrete nature of galaxies, and the rest of the expression accounts for the deviation from Poissonity, controlled by a parameter $\beta$. The negative binomial in this parametrization tends to a Poisson distribution for $\beta\rightarrow\infty$. The negative binomial distribution has the convenient statistical properties of allowing the mean and the variance to differ, which is used here to account for the known overdispersion of cosmological tracers \citep[e.g.,][]{Vakili2017,Pellejero2020}. 

The tracer density is modeled via a deterministic bias consisting in the product of a power law and two exponential factors accounting for high-density and low-density threshold bias:
\begin{equation}
\braket{\rho_{\rm tr}}_{dV} = f_{\rm tr} \times (1+\delta)^\alpha \times \exp\left[\left(-\frac{1+\delta}{\rho_\epsilon}\right)^\epsilon\right] \times \exp\left[\left(-\frac{1+\delta}{\rho^\prime_\epsilon}\right)^{\epsilon^\prime}\right] 
\end{equation}
where $f_{\rm tr}$ is the normalization constrained by the number density of the reference simulation, and $\theta=\{\alpha,\rho_\epsilon,\rho^\prime_\epsilon,\epsilon,\epsilon^\prime\}$ are free parameters of the model. Notice that the best-fitting values for these parameters will depend on the hierarchical cosmic web classification, i.e. given a cell belonging to the environment $i$ of the $\mathcal{T}$-web and to the environment $j$ of the $\mathcal{D}$-web, then $\theta=\theta_{ij}=\{\alpha_{ij},\rho_{\epsilon,ij},\rho^\prime_{\epsilon,ij},\epsilon_{ij},\epsilon^\prime_{ij}\}$. 

The effect of modelling the different ingredients of the bias described above is illustrated in Figure 5 in \cite{ColomaNadal2024}. Therein the authors test: (i) a local model with no threshold bias (first row), (ii) a local model including threshold bias, (iii) a nonlocal model with threshold bias including $\mathcal{T}-$web, and (iv), a nonlocal model with threshold bias and both $\mathcal{T}-$web and $\mathcal{D}-$web. It can be clearly seen how the predicted summary statistics improves from top to bottom, achieving higher accuracy towards small scales in the power spectrum and passing from a completely biased bispectrum (top) to an accurate one (bottom). Herein, we find a similar behaviour for QSOs as for haloes (Favole et al., in prep.).

\subsection{Bias model: calibration} \label{sec:bias_fit}

We calibrate the parameters of the model presented in \S\ref{sec:bias_theory} as follows.

We first obtain a proxy for the \texttt{Abacus} DM field at the target redshift by evolving the \texttt{Abacus} initial conditions (ICs) down to the target redshift using \texttt{ALPT}. It is important to rely on the \texttt{ALPT} version of the DM field in our framework, because we need to calibrate the bias using a density field which is consistent with the one underlying the DM lightcones described in \S\ref{sec:lightcones}. The ICs were properly downsampled to the final cell resolution $l\sim 5.5 \, h^{-1} \, {\rm Mpc}$ --- corresponding to $360^3$ cells mesh in a volume $V=(2 \, h^{-1} \, {\rm Gpc})^3$ --- using a Fourier- space sharp-$k$ filter\footnote{We adopt the same downsampling procedure followed to generate the \texttt{AbacusSummit} realizations at different resolutions. We refer the reader to the \texttt{Abacus} website (\url{https://abacussummit.readthedocs.io/en/latest/data-products.html\#initial-conditions}) for the details.}. The chosen resolution is the same as the one used to generate the DM lightcones presented in \S\ref{sec:lightcones}. The achieved mass resolution of our approximate simulations is $\sim 4.6\times10^{13} \, h^{-1} \, {\rm M}_\odot$ , i.e. a factor $2\times \sim 10^4$ lower than the reference \texttt{Abacus} simulation.

Afterwards, we apply the bias model presented in \S\ref{sec:bias_theory} and determine the free parameters of the model. We keep free all the parameters described in \S\ref{sec:bias_theory}. This implies a large number of parameters, but it allows to have a general description of the bias and describe different galaxy formation and evolution processes in distinct cosmic web environment. While it is in principle possible to reduce the dimensionality of the parameter space by carefully inspecting region by region and fixing (or avoid using) some of the parameters, we prefer to keep them free at this stage, although the clustering is not sensitive to some of them in some regions. We have verified that this choice does not overfit the model to the data, by applying the bias model on different cosmic realization of the \texttt{Abacus} simulations, not used for calibration. We determine the values of the free parameters by jointly fitting the QSO number counts probability distribution and the three-dimensional power spectrum. In particular, we first perform a visual scan to find suitable initial guesses for the parameters. Then, we perform a $\chi^2$ minimization of a loss function given by the sum of the aforementioned summary statistics, assuming Poisson errors on the PDF and Gaussian uncertainties on the power spectrum \cite[see e.g.,][]{Vakili2017}. The global mimimum of the function is found by means of a genetic algorithm.\footnote{\url{https://pypi.org/project/geneticalgorithm/}.} We typically reproduce the monopole and quadrupole from the \texttt{Abacus} reference catalog with few percent accuracy deviations up to the Nyquist frequency. In particular, we will present the results of the full redshift-dependent fitting of the QSO clustering in a forthcoming paper  (Favole et al., in prep.), while we focus in this paper on its application to generate Quaia mock catalogs. As will be explained in what follows, one should keep in mind that the preliminary calibration based on the \texttt{Abacus} simulations is just a starting point, and that a further tuning is needed to match the observed Quaia clustering. 

As a final step related to the bias calibration, we fine-tune the model to reproduce the angular clustering of the Quaia catalog. In particular, this operation is performed after the application of the subgrid model described in \S\ref{sec:subgrid} and of the injection of Quaia observational effects described in \S\ref{sec:quaia_obs_eff}. In fact, the bias model presented above reproduces the clustering of the DESI QSO sample, which in principle does not correspond to the one from Quaia. This fine-tuning is performed by varying the power-law exponents $\alpha$ until the large-scale normalization of the angular clustering of the mocks matches the one from the data after having applied the full mock-production pipeline. For both the $G<20.5$ and the $G<20$ samples, we find that the $\alpha$ parameter value that best fits the DESI QSO clustering must be multiplied by a factor $\sim 2$ to match the Quaia clustering amplitude. This means that the two Quaia samples have a larger bias than the DESI sample, consistently with the fact that the former have much lower magnitude cuts ($G<20.5$ and $G<20$, respectively) than DESI ($R<23.0$ in $r$-band \cite{Chaussidon2023}), and hence, are less sensitive to fainter QSOs. We stress that we calibrate the bias by relying only on the angular power spectrum $C_\ell$ in Fourier space, and no specific tuning of the configuration space angular clustering nor 3D clustering has been performed. We will validate the two-point angular correlation function in \S\ref{sec:validation}, and leave the study of 3D clustering measurements for a forthcoming paper (Shiferaw et al., in prep.). Furthermore, we notice that we are interestied in generating mock QSO catalogs both for the $G<20.5$ and the $G<20$ samples. To do so, we leave the freedom to have different model parameters for the two samples, in such a way to fit independently their angular clustering.

\subsection{Positions assignment and redshift space distortions}\label{sec:subgrid}

Once the QSO number counts field has been produced, we proceed with the assignment of QSO positions. To this end, we need to properly handle the fact that the used mesh has a limited resolution (cell size $l\sim 5.5 \, h^{-1} \, {\rm Mpc}$). Previous similar approaches relied  on a simple random uniform sampling of positions within the cell \citep{Kitaura2015,Zhao2021}. Recently, \cite{ForeroSanchez2024} devised a subgrid positions assignment based on a physically-motivated collapsing procedure of cosmological tracers towards attractors, defined as the maxima of the curvature tensor field.

In this work, we adopt an intermediate approach, which performs the first initial assignment in a similar fashion as \cite{ForeroSanchez2024}, but do not perform additional collapsing operations, which are not needed due to the sparse nature of the investigated QSO sample. In particular, we randomly assign positions of QSO inside the cell in correspondence of existing DM particles positions. In the cases where the QSO number counts exceeds the number of existing DM particles inside the same cell, we adopt a random uniform sampling of QSO positions within the cell. However, this is typically not the case for QSO due to their low number density and occurs only a very tiny amount of times ($\ll 1\%$ of the cells). 

Subsequently, we map the particle positions from real to redshift space following the standard redshift space distortions formula \cite{Kaiser1987}:
\begin{equation} \label{eq:rsd}
    \vec{s} = \vec{r} + b_v\frac{(\vec{v}\cdot \hat{n})\hat{n}}{aH}
\end{equation}
where $\vec{s}$ and $\vec{r}$ denote redshift-space and real-space Eulerian coordinates respectively, $b_v$ is the (linear) velocity bias, $\vec{v}$ is the velocity defined at point $\vec{r}$, $\hat{n}$ is the line-of-sight unit vector, and $a$ and $H$ are the scale factor and the Hubble parameter at the studied redshift respectively. Notice that since we are working on the lightcone, the redshift will vary throughout the box and so will the values for $a$ and $H$, which we then compute cell by cell.
The velocity field $\vec{v}=\vec{v}_{\rm coh} + \vec{v}_{\rm disp}$ is modelled as the sum of a large-scale coherent flow accounting for the bulk velocity flow and a small-scale dispersion velocity term modeling quasi-virialized motions \citep[see e.g.,][for the original presentation of the method and various applications]{Kitaura2012a,Kitaura2012b,Hess2013,Kitaura2014,Kitaura2016,Bos2019}. In particular, $\vec{v}_{\rm coh}$ corresponds to the velocity field defined on the mesh and obtained with the \texttt{ALPT} approximation. The three components of the small-scale velocity vector $\vec{v}_{\rm disp}$ are randomly sampled from a Gaussian distribution $\mathcal{G}(\mu=0,\sigma)$  with zero mean and standard deviation proportional to the density in the same cell through the power law $\sigma=A(1+\delta)^\gamma$. Once the velocity vector $\vec{v}$ has been obtained, we apply an additional nonlinear operation $\vec{v}'=\vec{v}^\eta,~\eta>1$, to supply more power to the velocity field. This transform maps the velocity components $0<v<1$\footnote{Here the velocities are already expressed in units $h^{-1}$ Mpc.} to values closer to zero and enhances the high-velocity tails at $v>1$, mimicking the highly non-Gaussian behavior of the velocity field from a N-body simulation as compared to the more Gaussian velocities from \texttt{ALPT}. The velocity model defined this way has four free parameters $\theta_v=\{b_v,A,\gamma,\eta\}$. Furthermore, we make the RSD parameters dependent on the $\mathcal{T}$-web environment, to also allow the modeling of velocities to be nonlocal \citep[see e.g.,][for previous applications of the same concept to the Lyman-$\alpha$ forest]{Sinigaglia2024a,Sinigaglia2024b}. This modeling is found to be accurate enough to reproduce the redshift-space anistropic clustering of the \texttt{Abacus} QSO sample at $z=1.4$ (the snapshot at the closest redshift available to the median redshift of the Quaia sample) with maximum deviations of $\lesssim 3\%$ and $\lesssim 10 \%$ up to $k\sim 1.0 \, h \, {\rm Mpc}^{-1}$ and $k\sim 0.6 \, h \, {\rm Mpc}^{-1}$ in the power spectrum monopole and quadrupole, respectively (Favole et al., in prep.). 

This defines our `spectroscopic' QSOs mock catalog, including QSO positions and velocities in redshift space.


\begin{figure}
    \centering
    \vspace{-1.5cm}
    {\huge \bf G<20.5}
    \includegraphics[width=\textwidth]{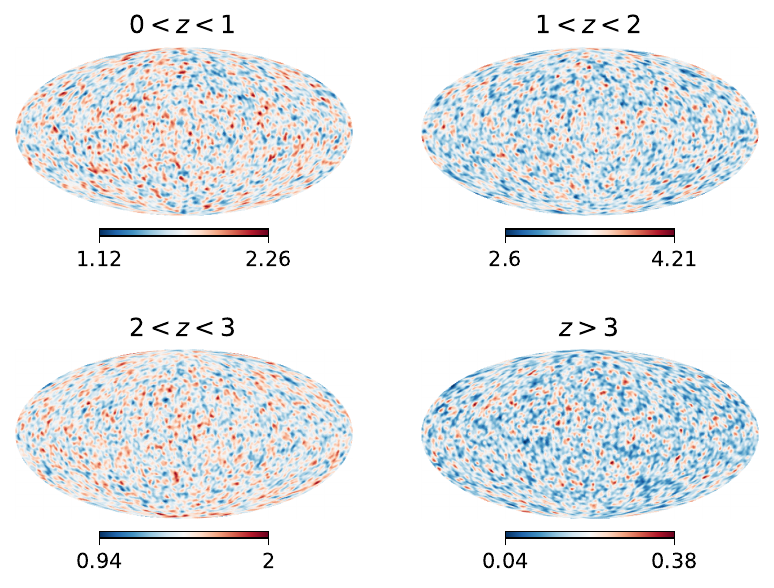}
    \caption{Full-sky maps of QSO number counts from one mocks realization (without angular selection function) and in different redshift bins, for the $G<20.5$ sample. The maps are color-coded by number counts per \texttt{Healpix} pixel.}
    \label{fig:maps_nosel_G20.5}
\end{figure}
\begin{figure}
    \centering
    \vspace{-1.5cm}
    {\huge \bf G<20}
    \includegraphics[width=\textwidth]{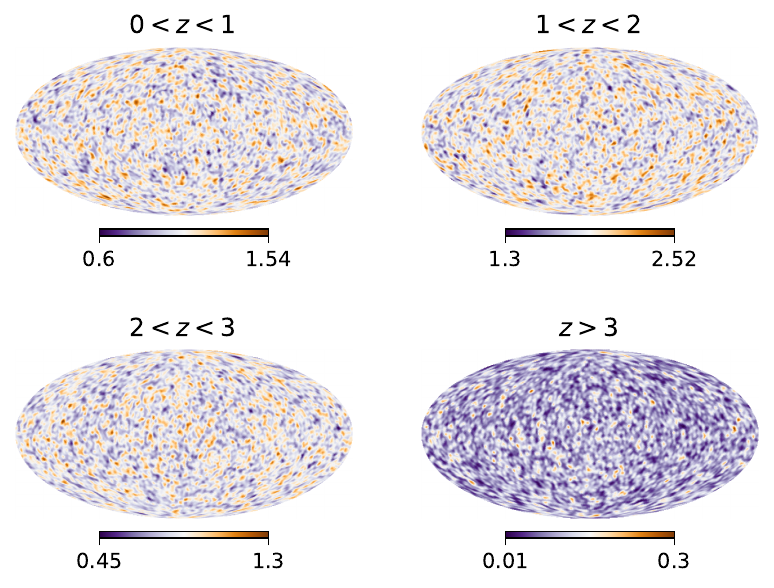}
    \caption{Same as Fig. \ref{fig:maps_nosel_G20.5}, but for the $G<20$ sample.}
    \label{fig:maps_nosel_G20}
\end{figure}

\subsection{Quaia observational effects}
\label{sec:quaia_obs_eff}

To finally obtain the Quaia-like catalogs from the  the spectroscopic mock catalogs described in the previous section, we inject into the catalog the observational features and systematics of the true Quaia data. In particular, we introduce:
\begin{itemize}
    \item redshift uncertainties, to mimic the spectrophotometric redshifts of Quaia: we first measure the redshift error as a function of redshift curve ($\Delta z$ vs $z$), and then apply it to each mock catalog. Then, once all QSOs have a redshift error assigned, we add to the spectroscopic redshift a random offset sampled from a Gaussian distribution with zero mean and standard deviation equal to the aforementioned redshift error;  
    \item the angular selection function: we use the Quaia selection function to extract only a subsample of QSO belonging to the same pixel\footnote{Throughout this work, we make use of the \texttt{Healpy} package and the related pixelization conventions.}. This is done by extracting only a random subset of the QSOs within the same pixel, with size proportional to the selection function value in that pixel. Specifically, we use the Quaia selection functions modeled for the low-$z$ and high-$z$ halves of the catalog separately, to account for some of the redshift dependence of the angular selection.
    \item the redshift distribution of the data: we measure the redshift distribution $n(z)$ from the data adopting $N=200$ redshift bins, and then subsample the spectroscopic catalog --- built on purpose to feature an overabundance of objects with respect to the Quaia catalog --- to match the observed $n(z)$ distribution. In each redshift bin, we model the uncertainty on the QSO number counts as a Poisson error.
\end{itemize}

After all these effects have been applied, the mocks catalogs contain all the relevant features present in the Quaia catalog.

\begin{figure*}
    \centering
    \vspace{-1.5cm}
    \includegraphics[width=\textwidth]{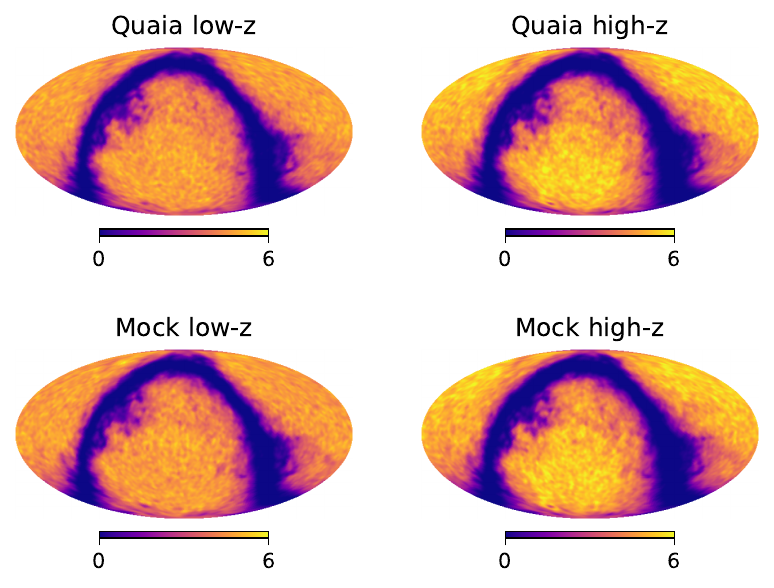}
    \caption{Full-sky maps of QSO number counts from the Quaia $G<20.5$ true catalog (top) and the respective mock catalog (bottom), showing the low-$z$ (left) and high-$z$ bins (right) separately. The maps are color-coded by number counts per \texttt{Healpix} pixel.}
    \label{fig:maps}
\end{figure*}

\begin{figure*}
    \centering
    \includegraphics[width=\textwidth]{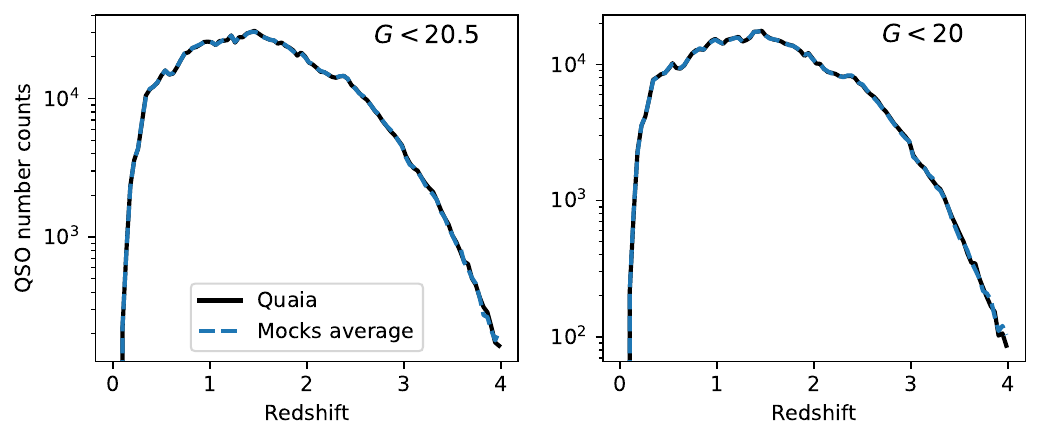}
    \caption{Redshift distributions of the quasars as a function of redshift for the $G<20.5$ (left) and $G<20$ (right) samples, respectively. The black solid line shows the mean $n(z)$ distribution from $100$ mocks, while the blue dashed shows the $n(z)$ distributions from the Quaia catalog.}
    \label{fig:z_distr}
\end{figure*}

\begin{figure*}
    \centering
    \includegraphics[width=\textwidth]{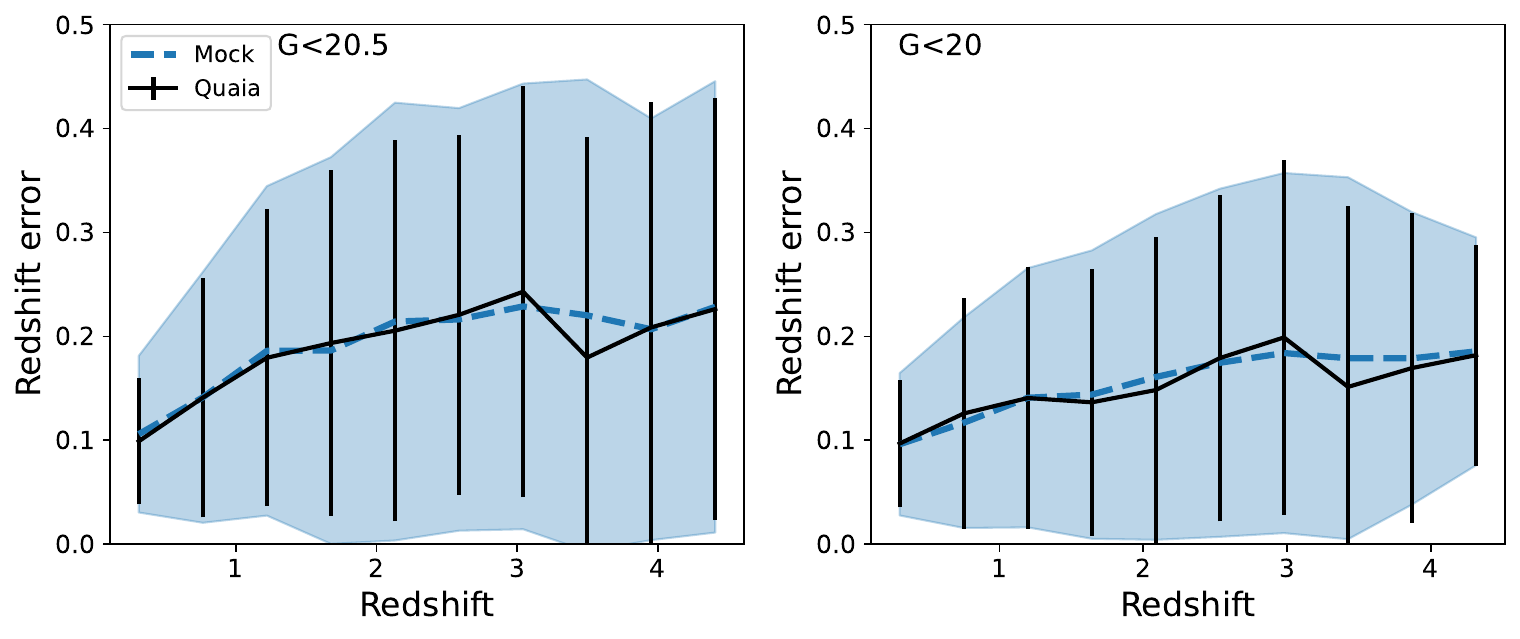}
    \caption{Redshift errors distribution as a function of redshift for the $G<20.5$ (left) and $G<20$ (right) samples, respectively. The black solid line shows the redshift errors distribution from the Quaia catalog, with uncertainties displayed as error bars. The blue dashed line shows the analogous distribution from one mock realization, with uncertainties displayed as an orange shaded region.}
    \label{fig:z_err}
\end{figure*}


\section{Validation of the mock catalogs}
\label{sec:validation}

In this section we present the validation of our mock catalogs against Quaia observational summary statistics.

Figs. \ref{fig:maps_nosel_G20.5} and \ref{fig:maps_nosel_G20} display full-sky maps of one mock realization without selection functions and in different redshift bins, for the $G<20.5$ and the $G<20$ samples, respectively. We notice that we can visually identify the cosmic web structure as traced by Quaia QSOs in all the maps. 

Fig. \ref{fig:maps} shows full-sky maps of QSO number counts from the Quaia true catalog (top) and mock catalog with the angular selection function applied (bottom), showing the low-$z$ (left) and high-$z$ bins (right) separately, for the $G<20.5$ sample. The maps are color-coded by number counts per \texttt{Healpix} pixel. A visual inspection reveals that the two maps feature a high degree of qualitative similarity, with the selection function clearly excluding the Galactic plane.

Fig. \ref{fig:z_distr} shows the redshift distributions of the QSOs in the mocks as compared to the Quaia catalog, for the $G<20.5$ (left) and the $G<20$ (right) subsamples. From a visual inspection of the resulting distributions, one notices that the redshift distribution from the mocks reproduce the one from the Quaia catalog with excellent agreement, as expected by construction from having sampled the former from the latter.

Fig. \ref{fig:z_err} displays the redshift errors distribution as a function of redshift for the $G<20.5$ (left) and $G<20$ (right) samples, respectively. The black solid line shows the average redshift error in each bin from the Quaia catalog, with standard deviation uncertainties displayed as error bars. The blue dashed line shows the analogous distribution from one mock realization, with uncertainties displayed as a blue shaded region. The mock catalog is found to reproduce the redshift uncertainties distribution as a function of redshift very well. In addition, we have explicitly verified that this results does not depend on the chosen specific mock realization and that all the mocks yield equivalently accurate results.

\begin{figure}
    \centering
    \includegraphics[width=\columnwidth]{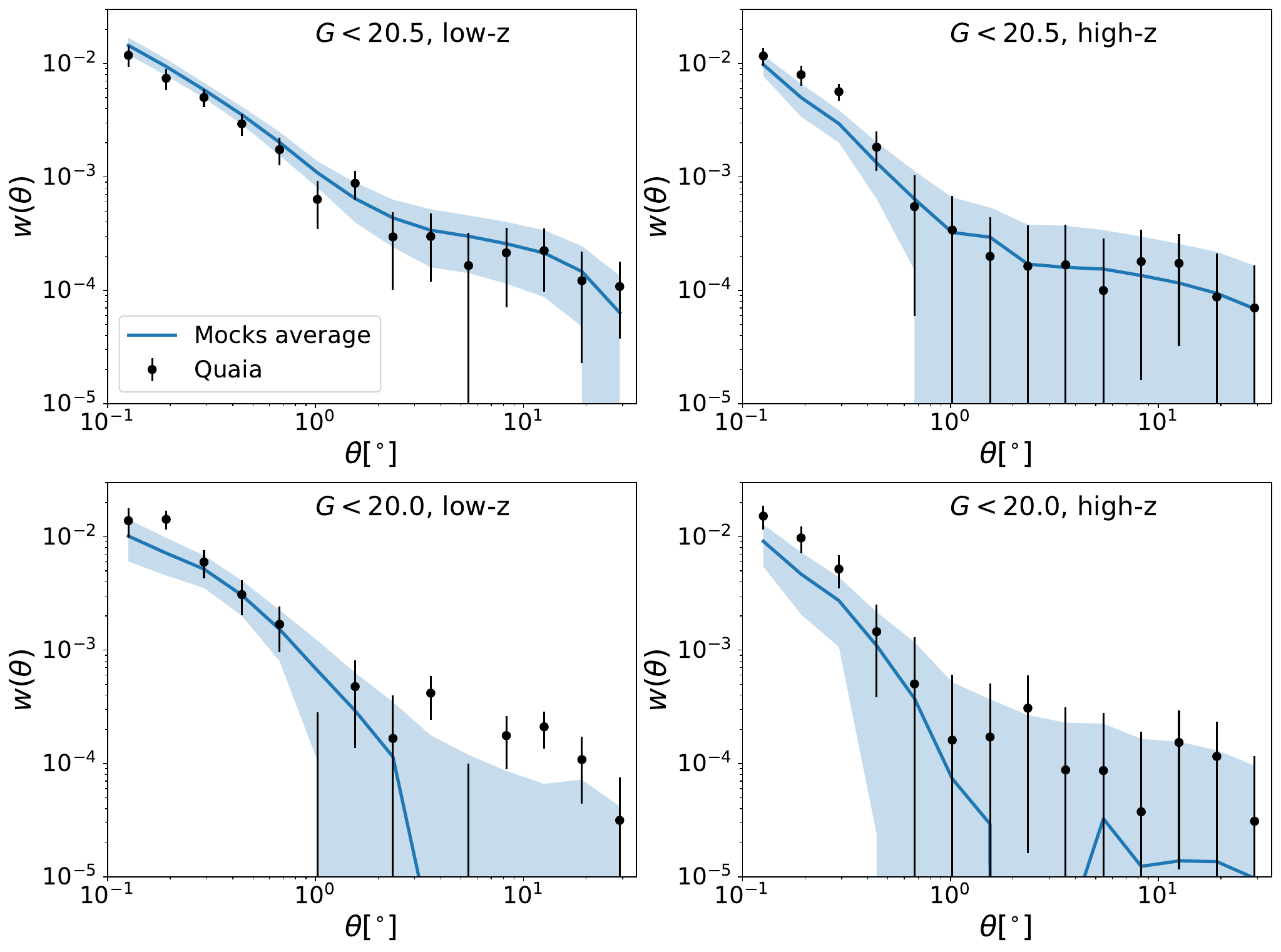}
    
    \caption{Angular two-point correlation function from Quaia (black dots) and the average of the mock catalogs (gray solid), measured in the following subsamples: low-$z$ at $G<20.5$ (top left), high-$z$ at $G<20.5$ (top right), low-$z$ at $G<20$ (bottom left), and high-$z$ at $G<20$ (bottom right). The gray shaded regions and the black error bars indicate the uncertainty on the mock and Quaia correlation functions, respectively, measured as variance from the $100$ mock realizations.}
    \label{fig:ang_2pcf}
\end{figure}

Then, we validate the angular clustering of the mocks comparing it to the one measured from Quaia data. We first look at the angular clustering in configuration space. We compute the angular two-point correlation function using the \texttt{Corrfunc} package \citep{Sinha2020}, creating random catalogs with $25$ times the number of objects in the true catalog. Figure \ref{fig:ang_2pcf} shows the angular two-point correlation function measured from the following subsamples: low-$z$ at $G<20.5$ (top left), high-$z$ at $G<20.5$ (top right), low-$z$ at $G<20$ (bottom left), and high-$z$ at $G<20$ (bottom right). The gray shaded regions and the black error bars indicate the uncertainty on the mock and Quaia correlation functions, respectively, measured as variance from the $100$ mock realizations. We find an excellent agreement over the whole probed range on angular scales for all the subsamples. In particular, the average from the $100$ mock catalogs fits well the data, achieving a compatibility within uncertainties in almost all the bins. A small deviation ($<2\sigma$) is found at $\theta\lesssim 0.3$ deg in the high-$z$ bin, both for the $G<20.5$ and $G<20$ samples, where the mocks systematically underpredict the Quaia clustering. This feature is the result of a lack of clustering amplitude on small scales. A more refined subgrid modelling would improve the agreement between the mocks and the data on these scales, but such scales will typically not be analyzed with Quaia. In particular, $\theta=0.3$ deg at $z\sim 4$ (the largest angular separation for which we find this deviation) corresponds to a distance of $\sim 25~h^{-1}~{\rm Mpc}$ comoving, smaller by a factor $\sim 2$ than the average QSO separation in the Quaia catalog. Therefore, we regard the achieved accuracy of our mocks as satisfying for the purpose of this work.

\begin{figure*}
    \centering
    \includegraphics[width=\textwidth]{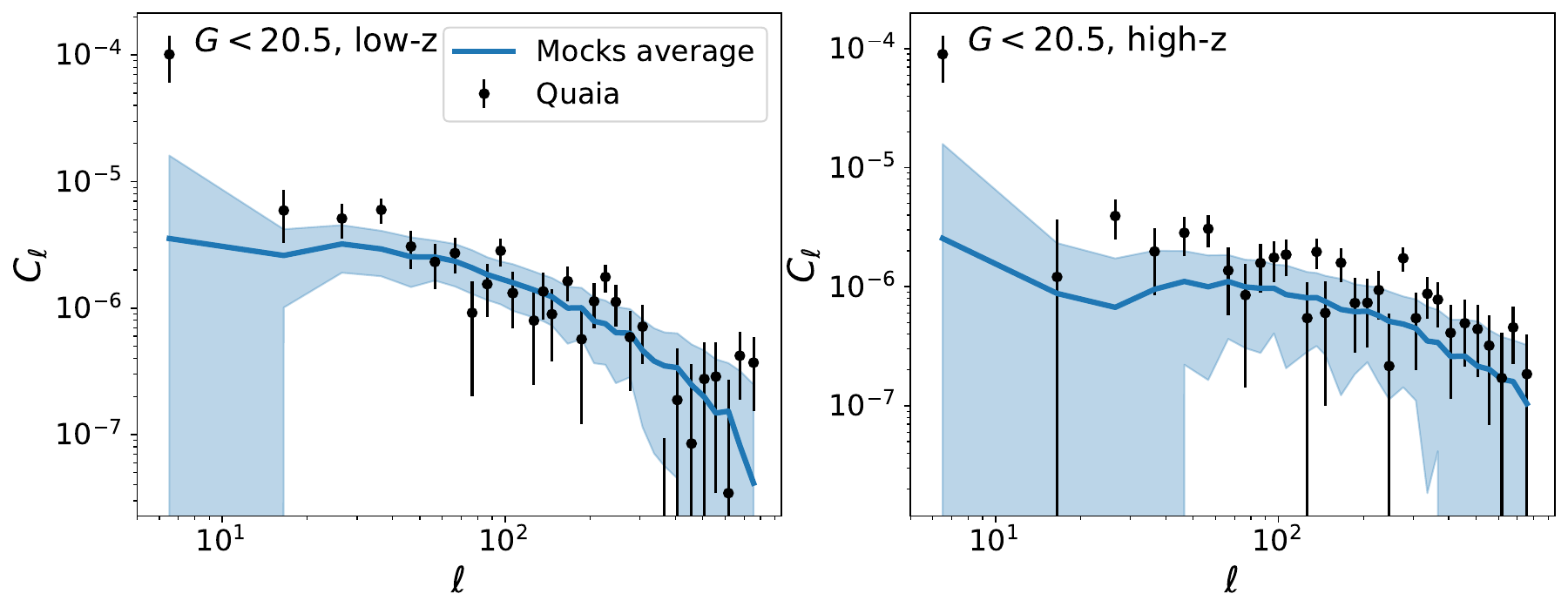}
    \includegraphics[width=\textwidth]{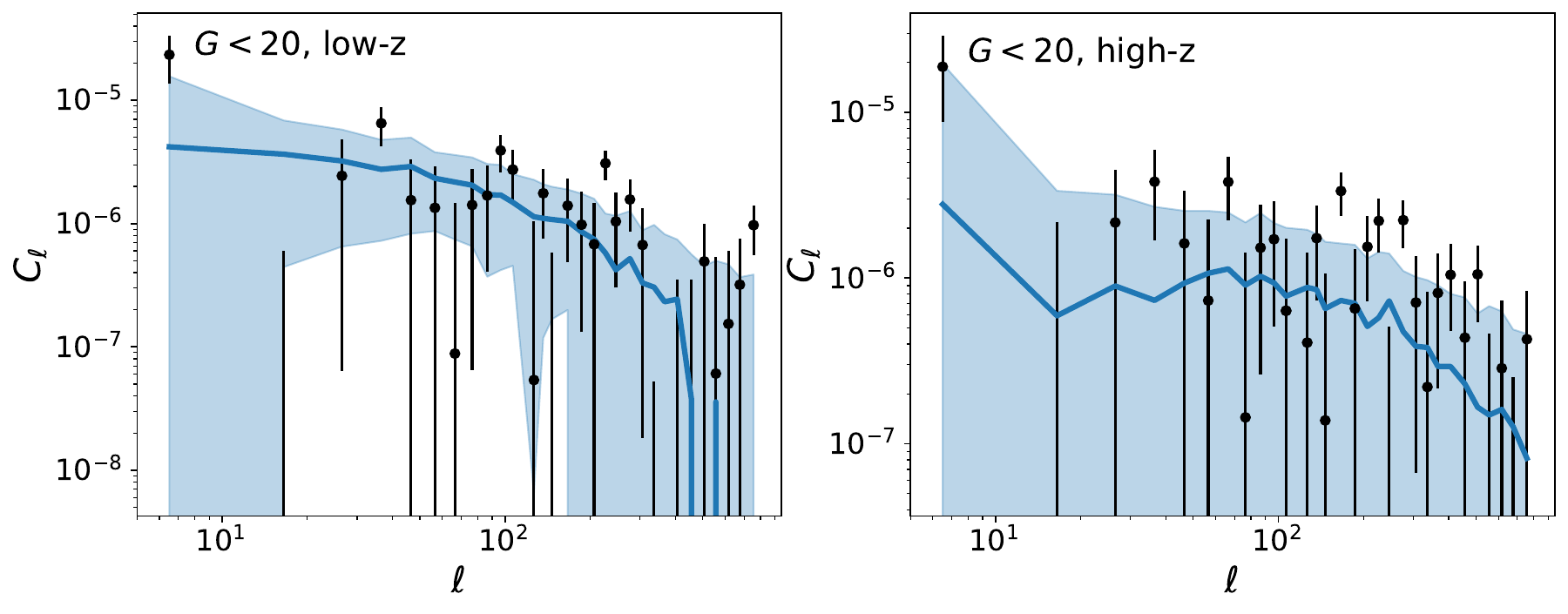}
    \caption{Angular power spectra from Quaia (black dots) and the average of the mock catalogs (blue solid), measured in the following subsamples: low-$z$ at $G<20.5$ (top left), high-$z$ at $G<20.5$ (top right), low-$z$ at $G<20$ (bottom left), and high-$z$ at $G<20$ (bottom right). The blue shaded regions indicate the variance of angular power spectrum measured from the $100$ mock realizations, while the black error bars indicate the uncertainties on the Quaia angular clustering computed via jackknife resampling \citep[see][]{StoreyFisher2024}.}
    \label{fig:c_ell}
\end{figure*}

We then compute the angular power spectra using the \texttt{NaMaster} package \cite{Alonso2019} and adopt the same setup and is the original Quaia analysis \cite{StoreyFisher2024}. Figure \ref{fig:c_ell} shows the measured angular power spectra, with the same structure and color scheme as in Figure \ref{fig:ang_2pcf}. This time, however, the black error bars indicate the uncertainties on the Quaia angular clustering computed via jackknife resampling \citep[see][]{StoreyFisher2024}. Similarly the Figure \ref{fig:ang_2pcf}, we find an excellent agreement between mocks and data for all the subsamples. We notice that the average $C_\ell$ from the mocks features a deviation from the Quaia $C_\ell$ in the lowest $\ell$ bin. However, it is worth stressing that such a data point in the Quaia $C_\ell$ is very high, which may be due to unmodelled systematics.

\begin{figure*}
    \centering
    \includegraphics[width=\textwidth]{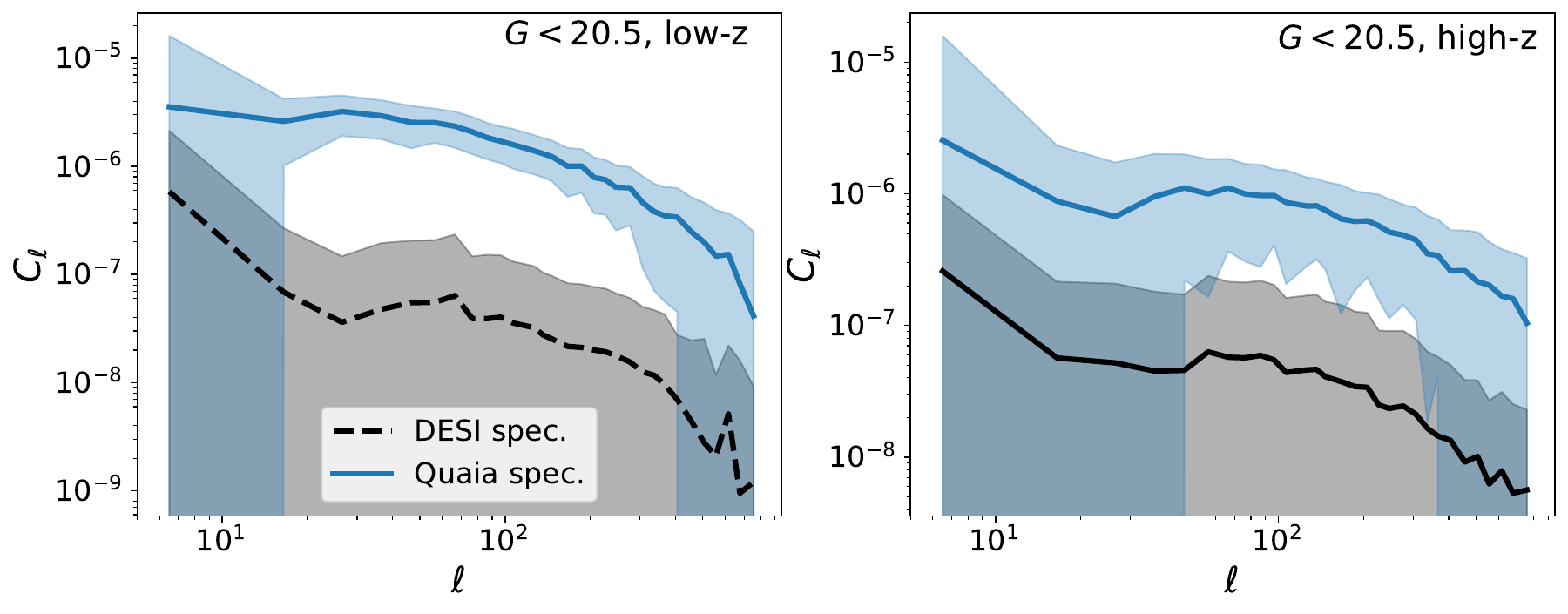}
    \caption{Angular power spectra from the average of $10$ DESI-like spectroscopic mock catalogs (black dashed) and the average of $10$ Quaia spectroscopic mock catalogs (blue solid), in the low-$z$ (left) and high-$z$ $G<20.5$ subsamples. The black and blue shaded regions indicate the variance of angular power spectrum measured from the $10$ DESI-like and Quaia mock realizations, respectively.}
    \label{fig:c_ell_desi}
\end{figure*}

To illustrate the impact of bridging the DESI EDR QSO sample to the Quaia sample through the final fine tuning described in \S\ref{sec:bias_theory}, in Figure \ref{fig:c_ell_desi} we show the comparison between the $C_\ell$ computed from the average of $10$ spectroscopic (systematics-free) DESI-like (before fine tuning, black dashed) and Quaia (after fine tuning, blue solid) mock catalogs, for the $G<20.5$ sample. In agreement with what reported in \S\ref{sec:bias_theory}, the Quaia sample features a higher bias with respect to the DESI-like sample, consistently with the fact that that DESI has a fainter magnitude cut.
 
\begin{figure*}
    \centering
    \includegraphics[width=\textwidth]{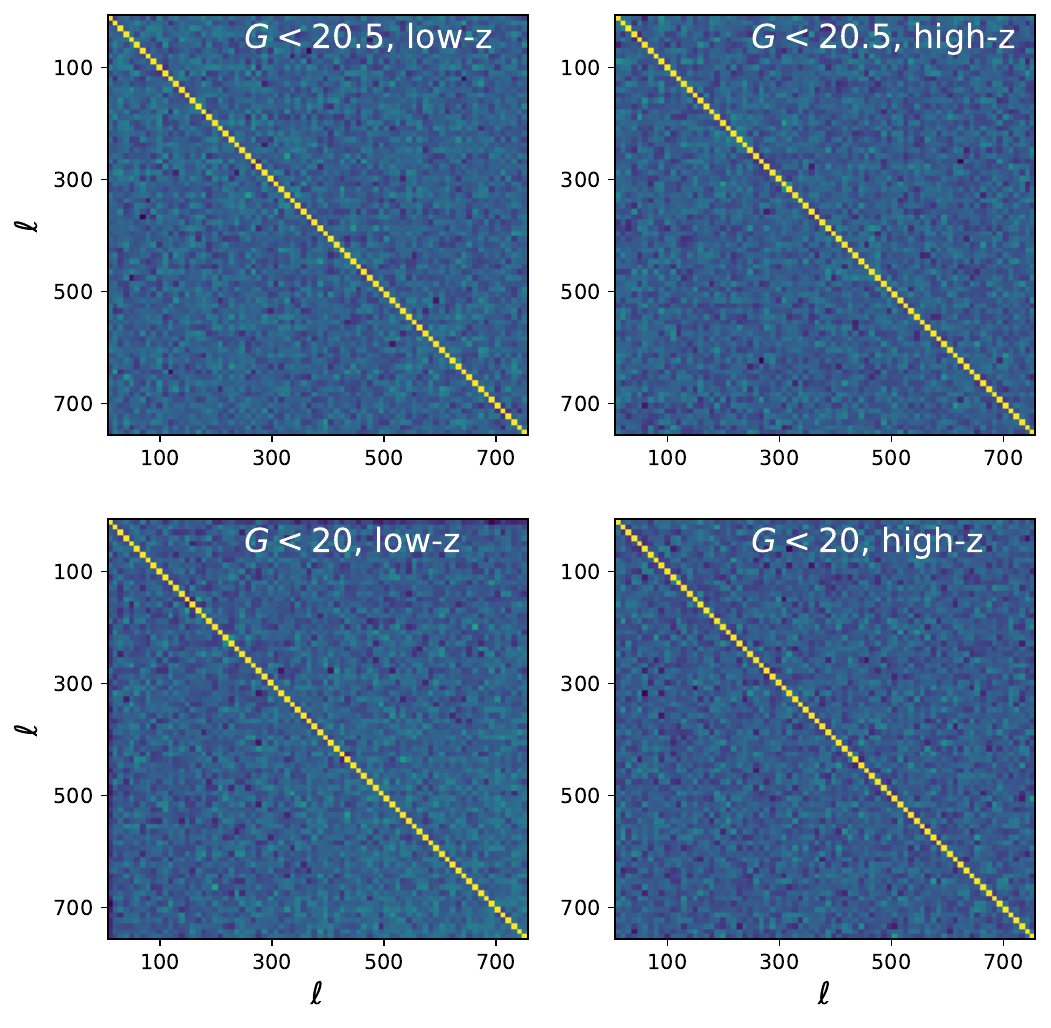}
    
    \caption{Normalized covariance matrices of the angular power spectra $C_\ell$ for the four studied subsamples: low-$z$ at $G<20.5$ (top left), high-$z$ at $G<20.5$ (top right), low-$z$ at $G<20$ (bottom left), and high-$z$ at $G<20$ (bottom right).}
    \label{fig:cov}
\end{figure*}

Figure \ref{fig:cov} shows the normalized covariance matrices of the angular power spectra $C_\ell$ as measured from ensemble of the mocks, for the four subsamples already shown in Figure \ref{fig:c_ell}: $G<20.5$ low-$z$ (top left), $G<20.5$ high-$z$ (top right), $G<20$ low-$z$ (bottom left), and $G<20$ high-$z$ (bottom right). The normalized covariance matrices are well-behaved: in particular, they are found to be positive-definite, with a clear correlation shown in the diagonal and with no major structures in the off-diagonal components. 

These tests prove that the QSO mock catalogs presented herein accurately reproduce the key summary statistics of the Quaia QSO catalog, and hence, they fulfill the requirements to be used to perform robust analysis of the real data. We notice that we have validated here only the angular clustering, both in configuration and in Fourier space. The reason behind this choice is that the analysis performed on Quaia data has so far been limited to these observables. In a  forthcoming paper, we will perform a comparison between the 3D clustering from Quaia data and from the mocks. However, we anticipate that preliminary checks that we have performed show good agreement, especially on large scales, relevant for BAO analysis.

\section{Impact of observational systematics}
\label{sec:systematics}

In this section, we test the impact of observational systematics on the measured clustering. To this end, we exploit the fact that our mock catalogs were constructed by gradually injecting the different systematics and we therefore have all the necessary information available. In particular, we test the impact of: (i) the angular mask, by comparing the clustering of the mocks with and without angular selection function, and (ii) redshift uncertainties, by comparing the clustering measured by relying on spectroscopic and on the spectrophotometric redshifts.   

\begin{figure*}
    \centering
    \includegraphics[width=\textwidth]{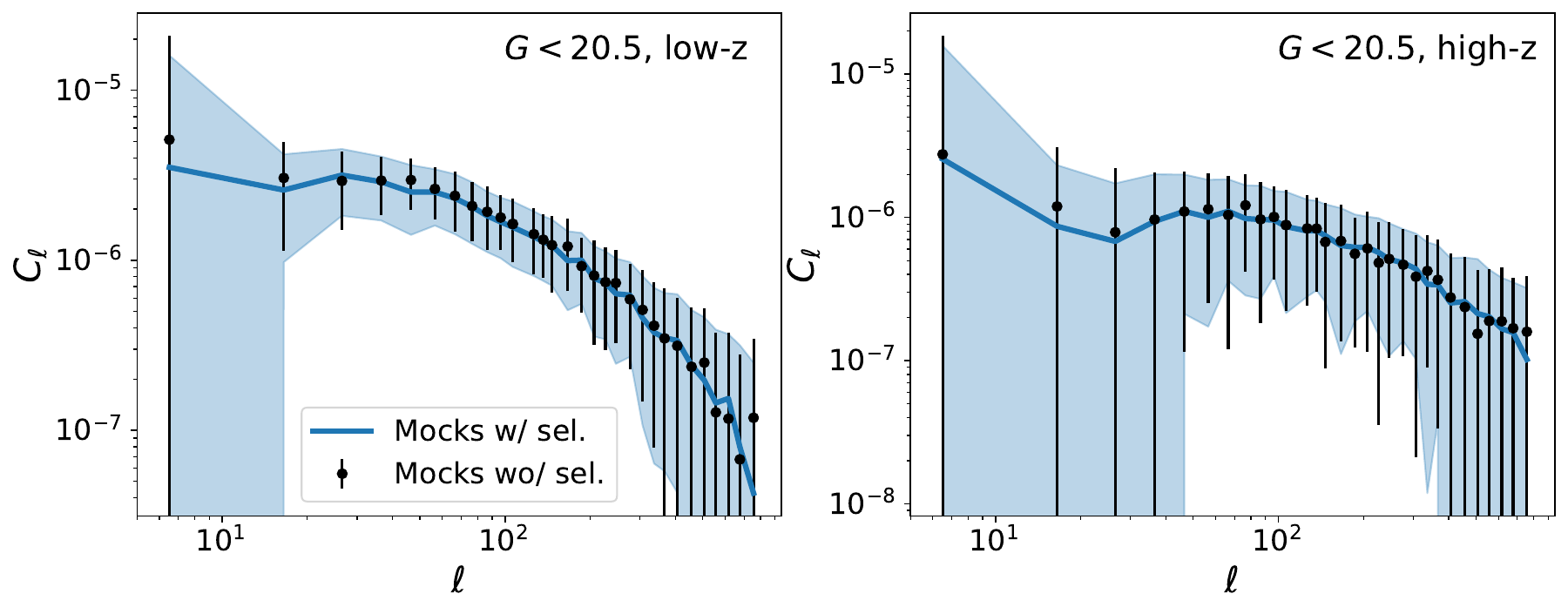}
    \includegraphics[width=\textwidth]{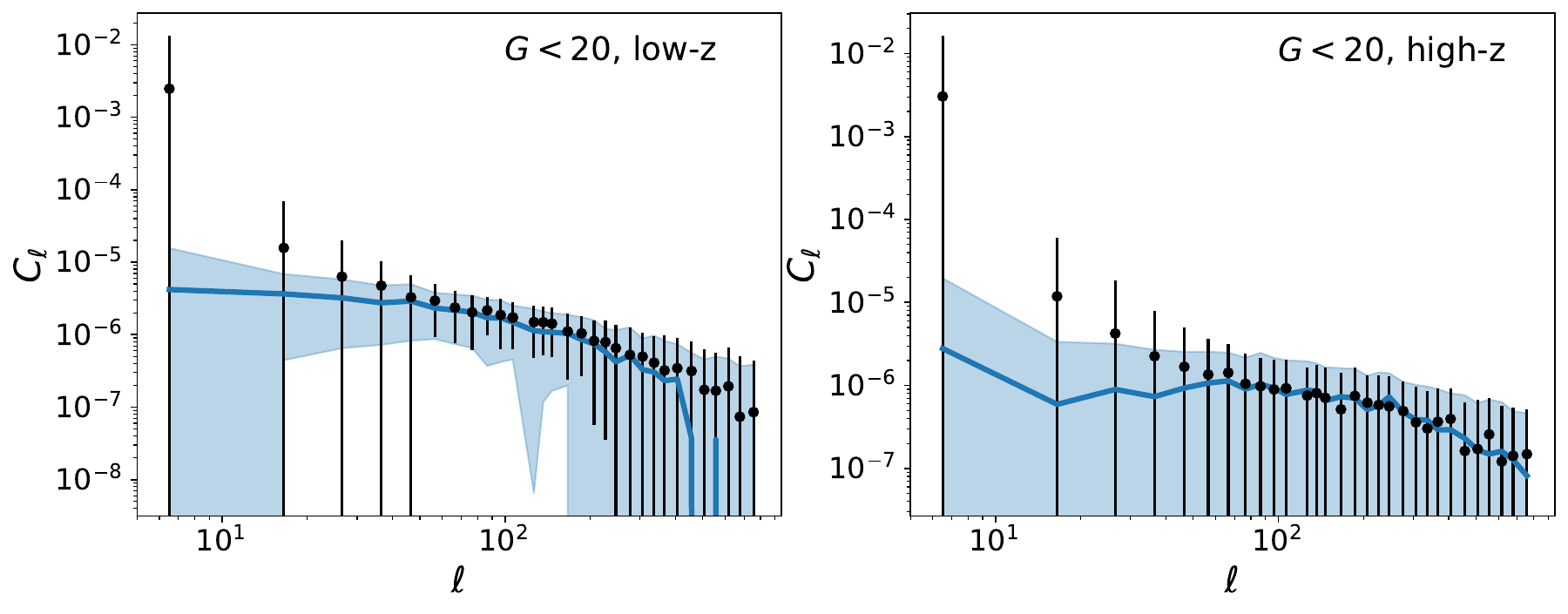}
    \caption{Angular power spectra from the average of $100$ mocks with (blue solid) and without angular selection function (black points), measured in the following subsamples: low-$z$ at $G<20.5$ (top left), high-$z$ at $G<20.5$ (top right), low-$z$ at $G<20$ (bottom left), and high-$z$ at $G<20$ (bottom right). The blue shaded regions indicate the variance of angular power spectrum measured from the mock realizations with angular selection, while the black error bars indicate the uncertainties on the angular clustering without angular selection.}
    \label{fig:c_ell_nosel}
\end{figure*}

Figure \ref{fig:c_ell_nosel} shows the measured angular power spectra displaying in blue the average from $100$ mocks including the angular selection function, and as black points the average from $100$ mocks without angular selection. The uncertainties were computed as standard deviation from the $100$ mocks in both cases. For the $G<20$ sample, we find an excellent agreement between mocks with and without angular selection function at $\ell\gtrsim 50$, while at $\ell\lesssim 50$ the average $C_\ell$ from the mocks without angular selection function shows a consistent excess compared to the $C_\ell$ from the mock with angular selection function, even though well within statistical uncertainties. For the $G<20.5$ sample, the $C_\ell$ are in excellent agreement for all the probed $\ell$s. From this comparison, we conclude that the lower $\ell$ should be treated with caution, as already mentioned above in the case of the comparison between data and mocks.

\begin{figure*}
    \centering
    \includegraphics[width=\textwidth]{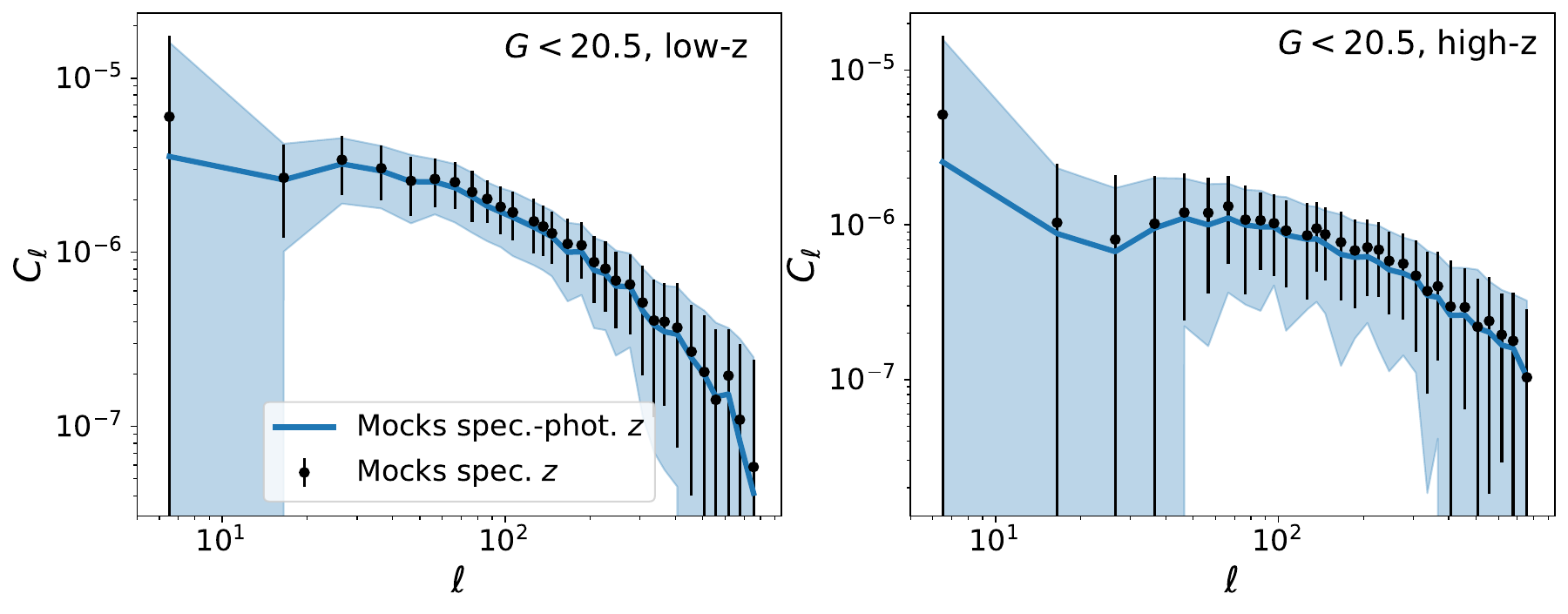}
    \includegraphics[width=\textwidth]{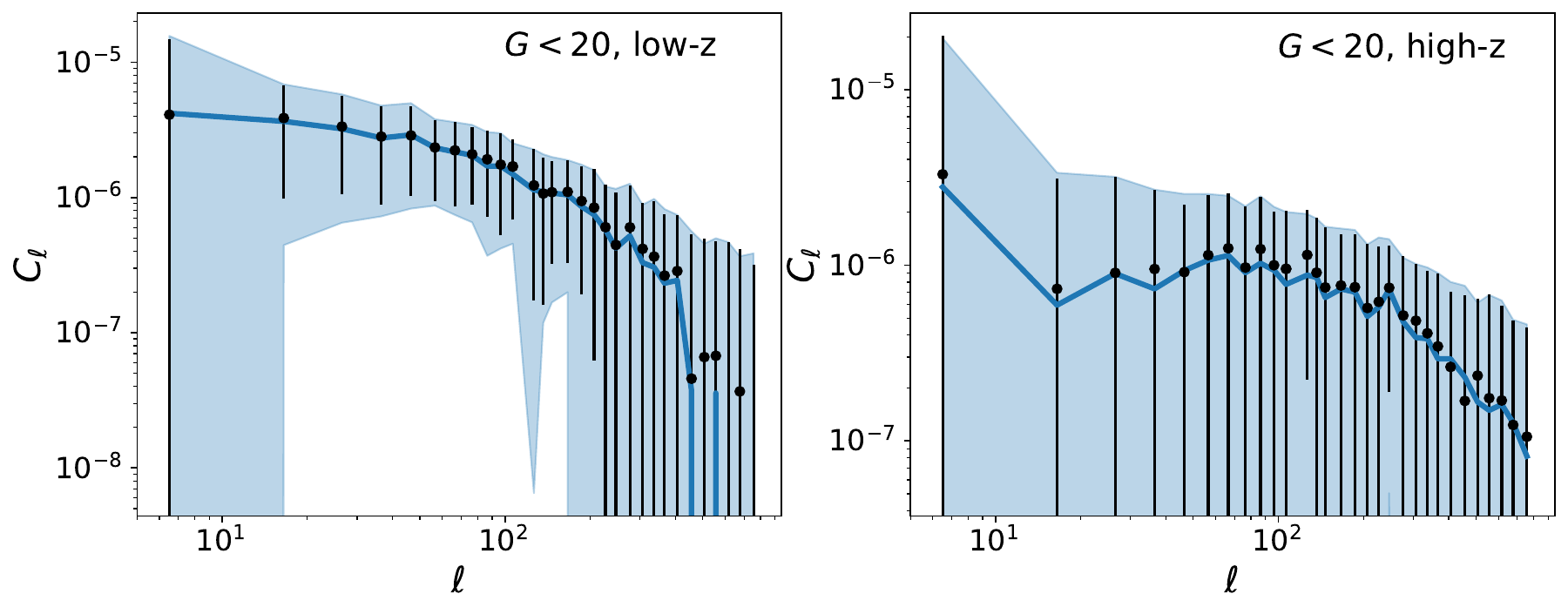}
    \caption{Angular power spectra from the average of $100$ mocks with Quaia-like spectrophotometric redshifts (blue solid) and with spectroscopic redshift (black points), measured in the following subsamples: low-$z$ at $G<20.5$ (top left), high-$z$ at $G<20.5$ (top right), low-$z$ at $G<20$ (bottom left), and high-$z$ at $G<20$ (bottom right). The blue shaded regions indicate the variance of angular power spectrum measured from the mock realizations with spectrophotometric redshifts, while the black error bars indicate the uncertainties on the angular clustering with spectroscopic redshifts.}
    \label{fig:c_ell_zspec}
\end{figure*}

Figure \ref{fig:c_ell_zspec} shows the measured angular power spectra displaying in blue the average from $100$ mocks relying on Quaia-like spectrophotometric redshift, and as black points the measurements using the spectroscopic redshifts. Also in this case, the uncertainties were computed as standard deviation from the $100$ mocks in both cases. For both the $G<20.5$ and $G<20.5$ samples we find that the measurements to be in excellent agreement and that the impact of the used redshifts is therefore negligible in the angular power spectra. We will address the impact of redshift errors in the 3D clustering in future work.  

\section{Conclusions}
\label{sec:conclusions}

In this work we present $100$ full-sky quasar spectrophotometric mock catalogue with lightcone geometry and smooth redshift evolution from $z=0$ to $z\sim 4$. These mock catalogs are tailored to reproduce the Gaia-unWISE Quasar Catalog (Quaia), and offer the possibility to robustly retrieve cosmological information from such a data set by providing a direct way to estimate covariance matrices, as well as serving as an ideal benchmark to test cosmological analysis methods. 

We start by generating `spectroscopic' QSO mock catalogs (i.e., with no redshift uncertainties). To do so, we generate real-space QSO number counts in cells by applying a hierarchical nonlinear nonlocal stochastic Eulerian bias model --- dubbed \texttt{Hicobian} \cite{ColomaNadal2024} --- to full-sky lightcone DM fields with smooth redshift evolution covering the redshift range $0<z\lesssim 4$ and produced through the \texttt{ALPT} \cite{Kitaura2013} on the lightcone within the \texttt{WebON} code (Kitaura \& Sinigaglia, in prep.). We calibrate the bias parameters upon high-fidelity QSO HOD catalogs constructed upon the \texttt{Abacus} N-body cosmological simulations suite \cite{Maksimova2021} and by fitting DESI EDR data \cite{Yuan2022,Yuan2024}. Afterwards, we assign QSO positions within the cell by relying on an educated assignment based on the existing \texttt{ALPT} DM particles positions. We model QSO velocities as a combination of large-scale coherent flows and small-scale quasi-virialized motions (modelling fingers-of-God) and apply a further power-law nonlinear transform to enhance the small-scale power of the velocity field. We determine the free parameters of this RSD model by fitting the monopole and quadrupole of the \texttt{Abacus} QSO HOD mocks. 
Once we have generated the spectroscopic mock catalogs, we turn them into realistic Quaia-like mock catalogs by injecting all the observational systematics: the spectrophotometric redshift uncertainties, the angular selection function, and the redshift $n(z)$ distribution. 

We assess the accuracy of the mocks by validating a variety of summary statistics: the full-sky QSO maps, the redshift uncertainty distributions as a function of redshift, the redshift $n(z)$ distribution, the angular power spectra and their normalized covariance matrices, and the angular two-point correlation function, finding excellent agreement for all of them. In particular, the high accuracy achieved in the reproduction of the angular clustering, as well as the well-behaved nature of the covariance matrices, supports the robustness of these catalogs for covariance matrix studies. In addition, we test the effect of the angular selection function and of redshift errors on the angular clustering, finding a negligible impact in most of the studied subsamples, with the exception of a deviation at $\ell\lesssim 50$ in the $G<20$ sample induced by the angular mask.

In future work, we will use these mocks to perform a variety of studies based on Quaia data: testing the impact of observational systematics such as the selection function and the spectrophotometric uncertainties, the assessment of the detectability of the BAO signal from Quaia, and the application of density field reconstruction techniques, among others. We publicly release the catalogs underlying this work, both with and without angular selection function. 

\section*{Data Availability}
The Quaia quasar catalogue and the corresponding selection function map are publicly available\footnote{\url{https://doi.org/10.5281/zenodo.10403370}} by their authors \citep{StoreyFisher2024}. The mock catalogs underlying this work are publicly available at \url{https://zenodo.org/records/19556442} ($G<20.5$) and at \url{https://zenodo.org/records/16789706} ($G<20$).

\section*{Structure of the mock catalogs}
The mock catalogs are released in \texttt{fits} table format, separately for the $G<20$ and the $G<20.5$ magnitude cuts. The table has the same structure as the Quaia true catalog, i.e. four columns corresponding to the three sky coordinates and the uncertainty on redshift: \texttt{ra}, \texttt{dec}, \texttt{redshift\_quaia}, \texttt{redshift\_quaia\_err}, all of them containing real double-precision floating-point numbers (\texttt{float64}). 


\acknowledgments
F.S., F.S.K., and G.F. acknowledge the Spanish Ministry of Science, Innovation and Universities (MICIU) for financing the \texttt{Big Data of the Cosmic Web} project: PID2020-120612GB-I00/AEI/10.13039/501100011033 under which this work was conceived and carried out, and the {\it Instituto de Astrofísica de Canarias} for support to the \texttt{Cosmology with LSS probes} project. This work and the mock catalogs generated herein are part of the \texttt{CosmicSignal} project (\url{http://www.cosmic-signal.org/}). F.S. acknowledges the support of the Swiss National Science Foundation (SNSF) 200021\_214990/1 grant. K.S.F. acknowledges the support of the Kavli Foundation. The authors acknowledge the CINECA award INAF\_C9A09, for the availability of high-performance computing resources and support. This research used resources of the National Energy Research Scientific Computing Center (NERSC), a DOE Office of Science User Facility supported by the Office of Science of the U.S. Department of Energy. The authors would like to thank the members of the Quaia team for useful comments and discussions, in particular Giulio Fabbian, András Kovács, and Carlos Hernández-Monteagudo.

\bibliographystyle{JHEP}
\bibliography{references.bib} 

\end{document}